\documentclass{aa} 

\usepackage[switch]{lineno} 

\usepackage{pdflscape} 

\usepackage{float}

\usepackage{bm}

\usepackage{adjustbox}
\usepackage{array}
\usepackage{graphicx}
\usepackage{enumitem}
\usepackage{siunitx}
\usepackage{multirow}
\usepackage{makecell}
\usepackage{upgreek}
\usepackage{soul,xcolor}
\setstcolor{red}
\usepackage{booktabs}
\usepackage[version=4]{mhchem}
\usepackage{svg}
\usepackage{comment}
\usepackage{natbib}
\bibpunct{(}{)}{;}{a}{}{,} 
\usepackage{txfonts}
\usepackage[colorlinks=true,
            linkcolor=blue,
            urlcolor=blue,
            citecolor=blue]{hyperref}
\makeatletter
\renewcommand*\aa@pageof{, page \thepage{} of \pageref*{LastPage}}
\makeatother

\makeatletter
\newcommand\thefontsize[1]{{#1 The current font size is: \f@size pt\par}}
\makeatother

\begin{document} 

\title{Astrometric exomoon detection by means of optical interferometry}

\titlerunning{Astrometric exomoon detection}
\authorrunning{T.~O.~Winterhalder et al.}

\author{T.\,O.~Winterhalder\inst{\ref{esog}}
        \and A.~M\'erand\inst{\ref{esog}}
        \and J.~Kammerer\inst{\ref{esog}}
        \and S.~Lacour\inst{\ref{lesia},\ref{esog}}
        \and M.~Nowak\inst{\ref{cam}}
        \and W.~O.~Balmer\inst{\ref{jhupa},\ref{stsci}}
        \and G.~Bourdarot\inst{\ref{mpe}}
        \and F.~Eisenhauer\inst{\ref{mpe}}
        \and A.~Glindemann\inst{\ref{esog}}
        \and S.~Grant\inst{\ref{mpe}}
        \and Th.~Henning\inst{\ref{mpia}}
        \and P.~Kervella\inst{\ref{lesia},\ref{french_chilean_lab}}
        \and G.-D.~Marleau\inst{\ref{mpia}, \ref{bernWP}, \ref{duisburg}}
        \and N.~Pourr\'e\inst{\ref{ipag}}
        \and E.~Rickman\inst{\ref{esa}}
        }

\institute{
    European Southern Observatory, Karl-Schwarzschildstrasse 2, D-85748 Garching bei München, Germany\label{esog} \and
    LIRA, Observatoire de Paris, Universit\'e PSL, CNRS, Sorbonne Universit\'e, Universit\'e de Paris, 5 place Jules Janssen, 92195 Meudon, France\label{lesia} \and
    Institute of Astronomy, University of Cambridge, Madingley Road, Cambridge CB3 0HA, United Kingdom\label{cam} \and
    Department of Physics \& Astronomy, Johns Hopkins University, 3400 N. Charles Street, Baltimore, MD 21218, USA\label{jhupa} \and
    Space Telescope Science Institute, 3700 San Martin Drive, Baltimore, MD 21218, USA\label{stsci} \and
    Max Planck Institute for extraterrestrial Physics, Giessenbachstra\ss e~1, 85748 Garching, Germany\label{mpe} \and
    Max Planck Institute for Astronomy, K\"onigstuhl 17, 69117 Heidelberg, Germany\label{mpia} \and
    French-Chilean Laboratory for Astronomy, IRL 3386, CNRS and U. de Chile, Casilla 36-D, Santiago, Chile\label{french_chilean_lab} \and
    Division of Space Research \&\ Planetary Sciences, Physics Institute, University of Bern, Gesellschaftsstr.~6, 3012 Bern, Switzerland\label{bernWP} \and
    Fakult\"{a}t f\"{u}r Physik, Universit\"{a}t Duisburg-Essen, Lotharstraße 1, 47057 Duisburg, Germany\label{duisburg} \and
    Univ. Grenoble Alpes, CNRS, IPAG, 38000 Grenoble, France\label{ipag} \and
    European Space Agency (ESA), ESA Office, Space Telescope Science Institute, 3700 San Martin Drive, Baltimore, MD 21218, USA\label{esa}
    }

   \date{Received 31 July 2025 / Accepted 5 September 2025}


   \abstract
   {With no conclusive detection to date, the search for exomoons, satellites of planets orbiting other stars, remains a formidable challenge. Detecting these objects, compiling a population-level sample and constraining their occurrence will inform planet and moon formation models and shed light on moon habitability.}
   {Here, we demonstrate the possibility of a moon search based on astrometric time series data, repeated measurements of the position of a given planet relative to its host star. The perturbing influence of an orbiting moon induces a potentially detectable planetary reflex motion.}
   {Based on an analytical description of the astrometric signal amplitude, we place the expected signatures of putative moons around real exoplanets into context with our current and future astrometric measurement precision.
   Modelling the orbital perturbation as a function of time, we then simulate the detection process given different target system configurations, instrumental measurement precisions and numbers of observational epochs to obtain the first astrometric exomoon sensitivity curves.
   }
   {
   The astrometric technique already allows for the detection and characterisation of favourable moons around giant exoplanets and brown dwarfs. Since the detection sensitivity of this method is mainly governed by the achievable astrometric precision, long-baseline interferometry lends itself ideally to this pursuit.
   We find that, on the basis of 12 epochs obtained with VLTI/GRAVITY, it is already today possible to infer the presence of a \SI{0.14}{M_{Jup}} satellite at a separation of \SI{0.39}{AU} around AF~Lep~b.
   Future facilities offering better precision will refine our sensitivity in both moon mass and separation from the host planet by several orders of magnitude.}
   {The astrometric method of exomoon detection, especially when applied to interferometric observations, provides a promising avenue towards making the detection of these elusive worlds a reality and efficiently building a sample of confirmed objects.
   With a future facility that achieves an astrometric precision of \SI{1}{\micro as}, probing for Earth-like moons within the habitable zone of a given star will become a realistic proposition.}


   \maketitle

\section{Introduction}
\label{section_introduction}

With the number of confirmed detections growing steadily, the field of exoplanet research is in the process of branching out into a diverse set of subdisciplines, ranging from formation modelling to the atmospheric characterisation of individual objects.
Another such ongoing pursuit is the search for low-mass companions to exoplanets, bodies conventionally called exomoons\footnote{While a strict definition of the term is yet to be agreed upon, we will use the word ``exomoon'' to mean a body gravitationally bound to a given exoplanet. Depending on the mass ratio between the two components, the term ``binary planet'' may be more appropriate, however.}.
With the exception of Mercury and Venus, every planet (and even some dwarf planets) in our Solar System harbours at least one moon.
It is thus plausible that planets orbiting other stars are accompanied by their own satellites.
A successful detection of such an object in an exoplanetary system would be the first step towards assessing the prevalence of exomoons around different types of host planets and building a population-level moon sample, which can help calibrate our models of moon formation, whether via capture \citep{hansen2019formation}, impact \citep{barr2017formation} or in-situ within the circumplanetary disc (e.g.\,\citealt{canup2006common, cilibrasi2021an}; see \citealt{cugno2024midinfrared} for the detection of a disc surrounding GQ~Lup~B).
Such knowledge will provide vital clues that can inform formation and evolutionary models (e.g.\,\citealt{peale2015origin, batygin2025determination}), the selection of follow-up targets deemed potentially habitable \citep{martinezrodriguez2019exomoons, dobos2022target} and the interpretation of putative biosignature detections in the future \citep{rein2014some}.

The different detection methods devised and pursued since the search for exomoons began in the wake of the first exoplanet detections (see e.g.\,\citealt{sartoretti1999on}) include searching for anomalous features in planetary transit observations (e.g.\,\citealt{szabo2006possibility, simon2007determination, kipping2009transit}), identifying the transit signatures of moons themselves as they eclipse free-floating planets (e.g.\,\citealt{limbach2021on, limbach2024occurrence}), measuring planetary radial velocity (RV) modulations (e.g.\,\citealt{ruffio2023detecting, vanderburg2021first}), monitoring for microlensing events (e.g.\,\citealt{liebig2010detectability}), probing for spectroastrometric signals (e.g.\,\citealt{agol2015center, vanwoerkom2024spectroastrometric}), attempting to image them directly (e.g.\,\citealt{lazzoni2020search}) and hunting for so-called Exo-Io signatures around hot Jupiters \citep{oza2019sodium}.
So far, however, no bona fide detection has been made (candidates include \object{Kepler-1625~b-i} and \object{Kepler-1708~b-i}; \citealt{teachey2018evidence, heller2019alternative, kreidberg2019no, teachey2020loose, kipping2022exomoon, heller2024large}; see also \citealt{kenworthy2023planetary}).

More recent studies simulating the survival rate of moons around planets at different orbital separations from their hosts are suggestive of why there has not yet been a confirmed detection. The sensitivity of transit studies -- inherently biased to short orbital periods and hence small separations between host and planet -- appears to be limited to a parameter space where moons are prone to being ejected or tidally disrupted (see e.g.\,\citealt{dobos2021survival}). To put it another way, exomoons are unlikely to survive in orbit around the planets accessible to transit observations. On the other hand, these simulations suggest that moons around far-out, long period exoplanets of the kind accessible to direct imaging studies have a high likelihood of remaining in orbit for several \SI{}{Gyr}. Here, however, our current contrast and resolution capabilities limit the achievable sensitivity of direct detection attempts and other methods such as spectroastrometry or moon transit monitoring are still to yield a definitive detection.

A ``middle ground'' between the two extremes is to hunt for moons around planets orbiting their host stars every few hundred days. In this window of opportunity, we expect the survival rate to be large enough to make a discovery possible, while also possessing the means of performing such a detection in the first place. As we shall show in this paper, precisely monitoring a given planet's position over time can reveal deviations from the expected Keplerian trajectory, that is a higher-order epicyclic or ``wobble'' motion, induced by the presence of an orbiting moon. Such a detection would amount to applying the astrometric method of planet detection by monitoring the periodic movement of a star to a planet--moon pair. This technique represents a readily applicable strategy towards making the first extrasolar satellite detection a reality.
The convenient possibility of constraining the mass and orbital elements of the moon from the astrometric time-series renders this method especially compelling.

At the heart of this new drive towards dynamic moon detection lies the unprecedented astrometric precision of the near-infrared interferometer GRAVITY at ESO's Very Large Telescope Interferometer (VLTI; \citealt{gravity2017first}). With an astrometric precision of \SI{50}{\micro as} \citep{lacour2019first}, the instrument has already facilitated remarkable constraints on the dynamical masses and orbital solutions of a large sample of directly detected planets. Notably, GRAVITY enabled an independent  dynamical detection and characterisation of $\upbeta$~Pic~c based solely on its perturbations to the orbit of $\upbeta$~Pic~b \citep{lacour2021mass} and provided the key data set used to resolve the first known brown dwarf, Gliese~229~B, into two near equal-mass binary components \citep{xuan2024cool, whitebook2024discovery}.

Here, we present an analytical modelling effort that aims to determine our current astrometric exomoon detection sensitivity and explore how future interferometric instruments and facilities might aid the hunt for these elusive objects.
This paper is structured as follows: in Sect.~\ref{section_astrometric_signal} we outline how to calculate the astrometric signal amplitude of a moon and to simulate its variation as a function of time. Sect.~\ref{section_single_epoch_detection_significances} deals with current and future capabilities of astrometric exomoon detection on the basis of several exemplary systems and the case of a putative moon around $\upbeta$~Pic~b in particular.
The modelling of detection attempts based on multiple astrometric epochs is described in Sect.~\ref{section_modelling_multi_epoch_detection_attempts} before the derived sensitivity curves are presented in Sect.~\ref{section_sensitivity_curves}. We conclude our study in Sect.~\ref{section_conclusions}.

\section{The astrometric signal}
\label{section_astrometric_signal}

The concept of an astrometric exomoon detection essentially amounts to a re-application of the astrometric method of planet detection.
As we shall see, the main differences and additional challenges with respect to the planet detection science case are the circumstance that the observable perturbation amplitudes are orders of magnitude smaller and the fact that we are required to perform the position measurements on the planets themselves.
To aid future target selection and observation planning, it is therefore crucial to understand the behaviour of the observable as a function of different parameters. Here, we briefly introduce the astrometric signal amplitude and variability in time.

\subsection{Astrometric signal amplitude}
To assess the detectability of a given star--planet--moon configuration, we are interested in the planet's deviation from a Keplerian orbit around the host as a consequence of being orbited by a moon.
Projected onto the plane of the sky, one can picture this perturbation as a time-dependent offset from the expected planetary position.
While varying periodically in right ascension (RA) and declination (Dec) according to the orbital parameters of the planet--moon system, we are initially mostly interested in the maximum radial deviation, a quantity we shall call the astrometric signal amplitude $A$ and which we can estimate via
\begin{equation}
    \label{equation_astrometric_signal}
    A = \left( \frac{G}{4 \pi^2} \right)^{1/3} \left( \frac{P}{M_\mathrm{pl} + M_\mathrm{m}} \right)^{2/3} \frac{M_\mathrm{m}}{d}
    = \frac{a_\mathrm{m}}{d} \frac{M_\mathrm{m}}{M_\mathrm{pl} + M_\mathrm{m}},
\end{equation}
where $G$ is the gravitational constant, $P$ is the orbital period of the planet--moon orbit, $d$ is the distance of the target system and $M_\mathrm{pl}$ and $ M_\mathrm{m}$ are the masses of the planet and moon, respectively \citep{sahlmann2013astrometric, quirrenbach2010astrometric}. For the second equality, we used Kepler's third law to substitute $P$ with an expression that includes the semi-major axis of the moon, $a_\mathrm{m}$. In general, Equation~\ref{equation_astrometric_signal} is an approximation that holds only for circular moon orbits.

To reiterate, the quantity $A$ is an amplitude and is therefore only a snapshot of the deviant wobble movement exhibited by the planet. In this sense, it is comparable to the RV semi-amplitude, $K$.
This analogy to the RV method for exoplanet detection extends even further: when hunting for planets using the RV method, we do not rely on a single measurement. To sample a large fraction of the phase-folded RV curve, many observations often spanning several orbital periods are required. This ensures two things: (1) ideally, the full radial velocity amplitude is registered and (2) the large number of epochs increases the detection significance. The same is true for the astrometric method of exomoon detection: while we do not know at which moment in time the planet shows the largest deviation and the measurement precision of our current instrumentation may not suffice to make a significant detection based on only a small number of observations, an extensive time-series of multiple astrometric epochs will solve both of these problems.
We therefore require not just an estimate of the perturbation amplitude as provided by Equation~\ref{equation_astrometric_signal}, but also its behaviour in time. In the following, we shall briefly explain how this modulation can be computed analytically.

\subsection{Simulating the signal variation in time}
\label{subsection_orbital_parametrisation}

To investigate the behaviour of the orbital perturbation experienced by the planet as a function of different parameters and thus its detectability with current and future instruments in more detail, we need to analytically describe the time variation of the signal. To this end, we defined two models.

\subsubsection{The star--planet model}
\label{subsubsection_star_planet_model}
First, we implemented an analytical two-body orbit model that is capable of integrating the trajectories of the host star and planet around their mutual centre of mass as a function of nine orbital parameters, namely the system's parallax, $\varpi$, the stellar and planetary masses, $M_\mathrm{s}$ and $M_\mathrm{pl}$, respectively, the semi-major axis of the planet, $a_\mathrm{pl}$, its inclination, $i_\mathrm{pl}$, eccentricity, $e_\mathrm{pl}$, longitude of ascending node, $\Omega_\mathrm{pl}$, argument of periastron, $\omega_\mathrm{pl}$, and time of periastron passage, $t_\mathrm{peri,\,pl}$. In broad terms, this is an algorithm that converts a given time of observation to a mean anomaly, solves Kepler's equation iteratively, performs a series of three-dimensional space rotations and scales the resulting stellar position by the planet--star mass ratio to obtain the planetary position. Performed for a series of time steps, this procedure yields the orbital trajectory of the two bodies as projected onto the plane of the sky.
Clearly, this model forms the baseline case of our investigation: a system devoid of a moon exhibiting perfect Keplerian two-body orbits.

\subsubsection{The star--planet--moon model}
\label{subsubsection_star_planet_moon_model}
On top of the star--planet model, we needed to simulate the system's behaviour under the influence of a moon in orbit around the planet. To this end, we employed the same framework as above but extended it by seven additional orbital parameters: the moon mass, $M_\mathrm{m}$, semi-major axis, $a_\mathrm{m}$, inclination, $i_\mathrm{m}$, eccentricity, $e_\mathrm{m}$, longitude of ascending node, $\Omega_\mathrm{m}$, argument of periplanet, $\omega_\mathrm{m}$, and time of periplanet passage, $t_\mathrm{peri,\,m}$. Thus, the star--planet--moon model incorporates 16 free parameters.

Whereas in the star--planet model we assumed the dynamical mass of the companion object or system to be entirely locked in the planet, we are here supposing that it is actually the sum of the planet and moon masses.
Thus, the trajectory obtained from the star--planet model can be reused as the trajectory of the planet--moon barycentre. Superimposing the trajectories of the planet and moon in their own reference frame onto this barycentric trajectory yields the space motions of the planet and moon around the barycentre of the entire system.
While this nested strategy avoids cumbersome three-body numerical integrations, it neglects higher-order gravitational interaction terms between star, planet and moon. Such a simplification appears reasonable, however, given that we intend to apply this model to stable systems exhibiting wide planetary orbits and for only negligible timescales compared to long-term orbital disruption and ejection processes. As \citet{kipping2010how} points out, this approach is justified provided the considered semi-major axis of the moon is smaller than \SI{0.531}{} times the planet's Hill radius, $R_\mathrm{Hill,\,pl}$. The latter is defined as
\begin{equation}
    \label{equation_hill_radius}
    R_\mathrm{Hill,\,pl} = a_\mathrm{pl} \left( \frac{M_\mathrm{pl}}{3M_\mathrm{s}} \right)^{1/3} .
\end{equation}

\subsubsection{Extracting the wobble}
\begin{figure*}
    \centering
    \includegraphics[width=0.99\textwidth]{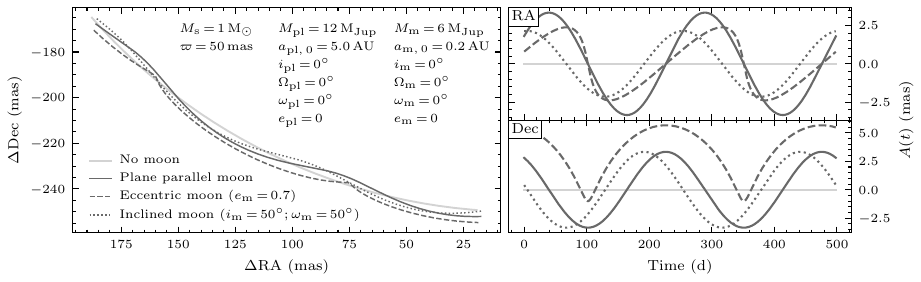}
    \caption{Orbital wobble caused by the perturbing gravitational influence of an exomoon. The \textit{left} panel shows the orbital trajectory of the moon-hosting planet relative to the star at the origin for different configurations, the \textit{right} panels display the corresponding astrometric signals in right ascension and declination as a function of time.
    For visualisation purposes, the mass ratio between moon and planet, $M_\mathrm{m}/M_\mathrm{pl}$, used in these examples is unrealistically high to the point that the system would more fittingly be described as a ``binary planet''.
    The plane parallel moon case corresponds exactly to the configuration given in the top right of the left hand panel. The eccentric and inclined examples were computed using the adjusted orbital elements specified in the brackets in the bottom left of the panel.
    The ``0'' in the indices of the semi-major axes indicate that these values correspond to the separation between the two bodies, that is between the planet--moon barycentre and the host star as well as between the moon and the planet, respectively. In other words, they are the sums of the semi-major axes of the respective hosts and orbiting bodies around their mutual centre of mass.
    Since we define the inclination of the moon to be \SI{0}{^\circ}
    and \SI{90}{^\circ} for face-on and edge-on orbits as projected onto the plane of the sky, respectively, simply increasing the inclination results in a dampening and eventual complete suppression of the astrometric signal in the declination component while -- as is evident in the bottom right panel -- the right ascension component remains unaffected. For the inclined example, we adjusted the argument of periastron to add a phase shift in the signal so as to better distinguish the different cases.
    }
    \label{figure_wobble_schematic}
\end{figure*}

Comparing the trajectories simulated using the star--planet and the star--planet--moon models yields the orbital deviation resulting from the presence of a moon. This perturbation can be visualised by plotting the orbital trajectories of the planet for both cases. The left hand panel of Fig.~\ref{figure_wobble_schematic} shows the resulting planetary wobble for a fictitious example system.
The shape of the wobble depends on the system configuration and can be deconstructed into RA and Dec components presented in the right hand panels of Fig.~\ref{figure_wobble_schematic}. Unlike the simplified description of the astrometric signal amplitude given in Equation~\ref{equation_astrometric_signal}, the star--planet and star--planet--moon models that underpin the parametrisation of the visualised wobble are capable of handling eccentric planet and moon orbits.

\section{Expected signals in the context of measurement uncertainties}
\label{section_single_epoch_detection_significances}

While, in general, an exomoon cannot be detected by a single astrometric measurement, it can be instructive to compare to our current astrometric measurement capabilities the theoretical perturbation in planet position as a consequence of an orbiting body.
To illustrate this, Table~\ref{table_astrom_amplitude_and_precision_examples} lists multiple known exoplanets and the astrometric signal amplitudes they would present for a series of assumptive companion moons.
Here, we have varied both the mass and semi-major axis of the moon. While the former assumes three distinct values for each case, for the reasons outlined in Sect.~\ref{subsubsection_star_planet_moon_model}, the latter is coupled to the respective Hill radius of the planet, $R_\mathrm{Hill,\,pl}$.
Apart from ensuring the applicability of the nested two-body framework we employed as an approximation of the three-body situation we are in reality confronted with, restricting the semi-major axis of the moon to values below a constant critical fraction, $\chi_\mathrm{crit}$, of the respective Hill radius also safeguards against modelling system configurations that are dynamically unstable.
Amongst other things, $\chi_\mathrm{crit}$ depends on the eccentricity and obliquity of the planetary orbit (e.g.\,\citealt{holman1999longterm}). Different estimates of the critical distance fraction include $\chi_\mathrm{crit} \approx \SI{0.36}{}$ and $\chi_\mathrm{crit} \approx \SI{0.49}{}$ for prograde \citep{barnes2002stability, domingos2006stable} and $\chi_\mathrm{crit} \approx \SI{0.93}{}$ for retrograde orbits \citep{domingos2006stable}.
To meet both the two-body approximation and the dynamical stability conditions, we therefore chose the two representative values of $\chi = \SI{0.5}{}$ and $\SI{0.1}{}$ to compute the case-dependent semi-major axes according to $a_\mathrm{m}=\chi R_\mathrm{Hill,pl}$.

The demonstrated and predicted astrometric measurement uncertainties associated with a set of current and future instruments listed in the bottom of Table~\ref{table_astrom_amplitude_and_precision_examples} place the computed signal amplitudes for the different system configurations into context. At present, GRAVITY is the only instrument capable of robustly resolving the perturbations predicted for a large subset of cases presented in Table~\ref{table_astrom_amplitude_and_precision_examples}. Classical imaging instruments, on the other hand, cannot be expected to be capable of detecting any but the most extreme cases, where the moon-to-planet mass ratios are unlikely large.

\begin{table}[t]
    \centering
    \caption{Astrometric signal amplitudes for different case examples in the \textit{top} and the approximate astrometric measurement precision of different current and future instruments in the \textit{bottom}.}
    \label{table_astrom_amplitude_and_precision_examples}
    \resizebox{\columnwidth}{!}{%
    \begin{tabular}{ccccc}
    \toprule
    Planet &
    System parameters &
    $M_\mathrm{m}$ (\SI{}{M_{Jup}}) &
    $\chi$ &
    $A$ (\SI{}{\micro as}) \\
    \midrule
    \midrule
    \multirow{6}{*}{\object{$\upbeta$~Pic~b}} &
    \multirow{6}{*}{\makecell[l]{
    $\varpi = \SI{50.93(15)}{mas}$~\hyperlink{ref_parallax}{\textsuperscript{1}} \\[1.9pt]
    $M_\mathrm{s} = \SI{1.83 (4)}{M_\odot}$~\hyperlink{ref_betapicb_brandt}{\textsuperscript{2}} \\[1.9pt]
    $M_\mathrm{pl}$ = $9.3_{-2.5}^{+2.6}\,\mathrm{M}_\mathrm{Jup}$~\hyperlink{ref_betapicb_brandt}{\textsuperscript{2}} \\[1.9pt]
    $a_\mathrm{pl} = \SI{10.26 (10)}{AU}$~\hyperlink{ref_betapicb_brandt}{\textsuperscript{2}} \\[1.9pt]
    $R_\mathrm{Hill,\,pl} = \SI{1.20 (11)}{AU}$
    }} &
    \multirow{2}{*}{1.0} & \SI{0.5}{} & $3000\,\pm\,600$ \\
     & & & \SI{0.1}{} & $600\,\pm\,130$ \\
     & \multirow{2}{*}{} & \multirow{2}{*}{0.1} & \SI{0.5}{} & $330\,\pm\,70$ \\
     & & & \SI{0.1}{} & $65\,\pm\,15$ \\
     & \multirow{2}{*}{} & \multirow{2}{*}{0.01} & \SI{0.5}{} & $33\,\pm\,8$ \\
     & & & \SI{0.1}{} & $6.6\,\pm\,1.5$ \\
     
    \midrule
    \multirow{6}{*}{\object{HR~8799~d}} &
    \multirow{6}{*}{\makecell[l]{
    $\varpi = \SI{24.46(5)}{mas}$~\hyperlink{ref_parallax}{\textsuperscript{1}} \\[1.9pt]
    $M_\mathrm{s}$ = $1.51_{-0.24}^{+0.38}\,\mathrm{M}_\odot$~\hyperlink{ref_hr8799d_stellar_mass_and_sma}{\textsuperscript{3}} \\[1.9pt]
    $M_\mathrm{pl} = \SI{10(3)}{M_{Jup}}$~\hyperlink{ref_hr8799d_mass}{\textsuperscript{4}} \\[1.9pt]
    $a_\mathrm{pl} = \SI{26.97 (73)}{AU}$~\hyperlink{ref_hr8799d_stellar_mass_and_sma}{\textsuperscript{3}} \\[1.9pt]
    $R_\mathrm{Hill,\,pl} = \SI{3.5 (4)}{AU}$
    }} &
    \multirow{2}{*}{1.0} & \SI{0.5}{} & $3800\,\pm\,900$ \\
     & & & \SI{0.1}{} & $770\,\pm\,180$ \\
     & \multirow{2}{*}{} & \multirow{2}{*}{0.1} & \SI{0.5}{} & $420\,\pm\,110$ \\
     & & & \SI{0.1}{} & $80\,\pm\,20$ \\
     & \multirow{2}{*}{} & \multirow{2}{*}{0.01} & \SI{0.5}{} & $42\,\pm\,11$ \\
     & & & \SI{0.1}{} & $8\,\pm\,2$ \\
     
    \midrule
    \multirow{6}{*}{\object{HD~206893~c}} &
    \multirow{6}{*}{\makecell[l]{
    $\varpi = \SI{24.53(4)}{mas}$~\hyperlink{ref_parallax}{\textsuperscript{1}} \\[1.9pt]
    $M_\mathrm{s}$ = $1.32_{-0.05}^{+0.07}\,\mathrm{M}_\odot$~\hyperlink{ref_hd206893c_hinkley}{\textsuperscript{5}} \\[1.9pt]
    $M_\mathrm{pl}$ = $12.7_{-1.0}^{+1.2}\,\mathrm{M}_\mathrm{Jup}$~\hyperlink{ref_hd206893c_hinkley}{\textsuperscript{5}} \\[1.9pt]
    $a_\mathrm{pl}$ = $3.53_{-0.06}^{+0.08}\,\mathrm{AU}$~\hyperlink{ref_hd206893c_hinkley}{\textsuperscript{5}} \\[1.9pt]
    $R_\mathrm{Hill,\,pl} = \SI{0.51 (2)}{AU}$
    }} &
    \multirow{2}{*}{1.0} & \SI{0.5}{} & $460\,\pm\,30$ \\
     & & & \SI{0.1}{} & $92\,\pm\,7$ \\
     & \multirow{2}{*}{} & \multirow{2}{*}{0.1} & \SI{0.5}{} & $49\,\pm\,4$ \\
     & & & \SI{0.1}{} & $9.8\,\pm\,0.7$ \\
     & \multirow{2}{*}{} & \multirow{2}{*}{0.01} & \SI{0.5}{} & $5.0\,\pm\,0.4$ \\
     & & & \SI{0.1}{} & $0.99\,\pm\,0.07$ \\
    \bottomrule
    \end{tabular}
    }
    
    \vspace{1em}

    \resizebox{\columnwidth}{!}{%
    \begin{tabular}{ccccccc}
    \toprule
     &
     SPHERE &
     GPI &
     GRAVITY &
     PLANETES\hyperlink{ref_future_instrument}{\textsuperscript{*}} &
     MICADO\hyperlink{ref_micado_instrument}{\textsuperscript{\textdagger}} &
     KBI\hyperlink{ref_future_facility}{\textsuperscript{\textdagger\textdagger}} \\
    \midrule
    \midrule
    $\sigma_A$ (\SI{}{\micro as}) &
    \SI{1500}{}~\hyperlink{ref_sphere_accuracy}{\textsuperscript{6}} &
    \SI{1000}{}~\hyperlink{ref_gpi_accuracy}{\textsuperscript{7}} &
    \SI{50}{}~\hyperlink{ref_gravity_accuracy}{\textsuperscript{8}} &
    \SI{10}{}~\hyperlink{ref_priv_comm_accuracy}{\textsuperscript{9}} &
    \SI{400}{}~\hyperlink{ref_priv_comm_accuracy}{\textsuperscript{9}} &
    < \SI{1}{}~\hyperlink{ref_kbi_accuracy}{\textsuperscript{10}} \\
    \bottomrule
    \end{tabular}
    }
    \tablefoot{In the \textit{top}, $\varpi$, $M_\mathrm{s}$, $M_\mathrm{pl}$ and $a_\mathrm{s}$ are the system's parallax, stellar mass, planetary mass and semi-major axis, respectively. $R_\mathrm{Hill,\,pl}$ is the Hill radius that results from the system parameters according to Equation~\ref{equation_hill_radius} when propagating the uncertainties associated with the different quantities involved. The semi-major axis of the moon used to compute the signal amplitude, $A$, via Equation~\ref{equation_astrometric_signal}, is the respective Hill radius scaled by the constant fraction, $\chi$. Again, all uncertainties were propagated until arriving at the signal amplitude. In the \textit{bottom}, $\sigma_A$ is the approximate astrometric measurement uncertainty achievable with different instruments. \\
    \hypertarget{ref_future_instrument}{\textsuperscript{*}} Future VLTI instrument (see Sect.~\ref{section_single_epoch_detection_significances});
    \hypertarget{ref_micado_instrument}{\textsuperscript{\textdagger}} First-generation ELT instrument \citep{davies2021micado};
    \hypertarget{ref_future_facility}{\textsuperscript{\textdagger\textdagger}} Future kilometre-baseline interferometric facility \citep{bourdarot2024kilometer}; \\
    \hypertarget{ref_parallax}{\textsuperscript{1}} \citet{gaia2023dr3};
    \hypertarget{ref_betapicb_brandt}{\textsuperscript{2}} \citet{brandt2021precise};
    \hypertarget{ref_hr8799d_stellar_mass_and_sma}{\textsuperscript{3}} \citet{zurlo2016first};
    \hypertarget{ref_hr8799d_mass}{\textsuperscript{4}} \citet{marois2008direct};
    \hypertarget{ref_hd206893c_hinkley}{\textsuperscript{5}} \citet{hinkley2023direct};
    \hypertarget{ref_sphere_accuracy}{\textsuperscript{6}} \citet{maire2021lessons};
    \hypertarget{ref_gpi_accuracy}{\textsuperscript{7}} \citet{wang2016orbit};
    \hypertarget{ref_gravity_accuracy}{\textsuperscript{8}} \citet{lacour2019first};
    \hypertarget{ref_priv_comm_accuracy}{\textsuperscript{9}} priv. comm.;
    \hypertarget{ref_kbi_accuracy}{\textsuperscript{10}} \citet{bourdarot2024kilometer}
    }
\end{table}

Clearly, the system configurations envisaged in Table~\ref{table_astrom_amplitude_and_precision_examples} are entirely imaginative and serve to illustrate how the signal strength changes for a set of exemplary cases.
If one were to attempt an astrometric exomoon detection in earnest, however, one would ideally want to avoid having to resort to a blind search around an arbitrary planet. This is due to the prolonged period of time throughout which astrometric epochs need to be collected for in order to achieve a positive detection. Instead, an additional line of evidence that hints towards the potential existence of an orbiting body should be the starting point of an efficient astrometric moon hunt.
In this regard, the case of $\upbeta$~Pic~b warrants further attention. Recently, \citet{poon2024potential} have argued that the planet's potentially increased obliquity (to be confirmed by future observations) could be explained by the perturbing influence of an exomoon.
They also give estimates as to the moon mass and semi-major axis required to induce the observed planetary obliquity.
Hereafter, we use the shorthand ``P24-moon'' to refer to objects that comply with these constraints. Prior to the suggestion of the ``P24-moon'', \citet{macias2024constraining} have already ruled out moons more massive than \SI{0.25}{M_{Jup}}.

\begin{figure}
    \centering
    \includegraphics[width=0.99\columnwidth]{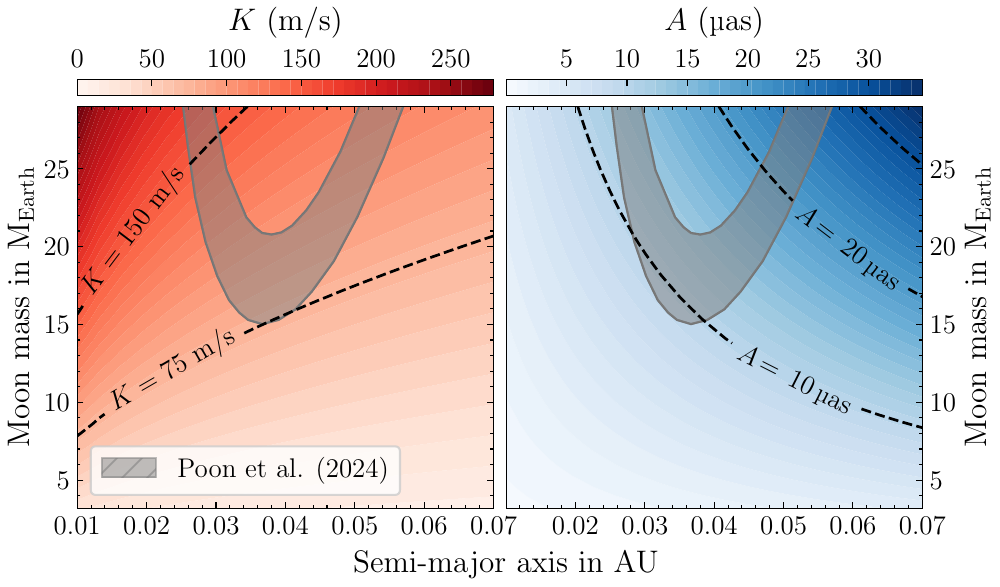}
    \caption{
    Signal amplitudes in the vicinity of the putative exomoon around $\upbeta$~Pic~b suggested by \citet{poon2024potential} as a function of moon mass and semi-major axis for RV studies on the \textit{left} and astrometric searches on the \textit{right}.
    The grey shaded region is the parameter space where a P24-moon is likely to reside if it is indeed real.
    The dashed lines indicate contours delineating the provided signal strengths.
    }
    \label{figure_RV_astrom_signal_amplitude_maps}
\end{figure}

To understand whether our current instrumentation is capable of picking up on the astrometric signal of such a P24-moon, we can compute the expected amplitude in the vicinity of the moon mass and semi-major axis constraints using the models defined in Sect.~\ref{subsection_orbital_parametrisation}. This region of the parameter space is visualised in the right panel of Fig.~\ref{figure_RV_astrom_signal_amplitude_maps}.
It demonstrates that a P24-moon would induce an astrometric wobble amplitude of \SI{10}{} to \SI{20}{\micro as}.
Despite its unprecedented astrometric precision of \SI{50}{\micro as}, such a perturbation would likely be difficult to detect using GRAVITY. 
However, plans are already underway to improve the astrometric accuracy of the existing GRAVITY instrument by addressing certain limitations \citep[e.g., optical aberrations in][]{2021A&A...647A..59G}, as well as by halving the operational wavelength -- observing in the Y- and J-bands instead of the K-band. These efforts are part of the VLTI/PLANETES technological development program at ESO\footnote{Funded by the European Research Council: \url{https://cordis.europa.eu/project/id/101142746/fr}}.
A future instrument developed under this program would be expected to achieve a precision of \SI{10}{\micro as} \citep[as predicted in][]{2014A&A...567A..75L}, providing the capability to search for a P24-moon around $\upbeta$~Pic~b.
Comparability of the expected signal amplitude with the instrumental measurement uncertainty implies a single-epoch signal-to-noise comparable to 1. In Sect.~\ref{section_modelling_multi_epoch_detection_attempts}, we shall see how the detection significance increases when acquiring multiple epochs.

The unique potential inherent to the astrometric method is reinforced by the comparison to the detection capabilities of other methods. For one thing, strategies that hinge on observations of the planet transiting the stellar disc are not applicable to this non-transiting planet. What is more, \citet{poon2024potential} point out that transits of the putative moon in front of the planet are unlikely. An alternative avenue towards eventual detection might lie in a comprehensive RV monitoring programme of the exoplanet.
To gauge whether such a strategy is feasible for detecting a P24-moon, we visualised the expected RV semi-amplitude, $K$, as a function of moon mass and semi-major axis in the left hand panel of Fig.~\ref{figure_RV_astrom_signal_amplitude_maps}. As expected, the sensitivity gradient of the RV method is opposite to that of the astrometric technique: while both signals increase with the moon mass, the astrometric signal is stronger for larger semi-major axes, whereas the RV semi-amplitude decreases with larger separations. To probe the entire parameter space region predicted for the P24-moon requires an RV precision of approximately \SI{75}{m \per s}.
Recent studies have already demonstrated measurements of planetary RVs to be possible when coupling an adaptive optics system with a high-resolution spectrograph.
\citet{denis2025characterisation} could constrain the RV of \object{AF~Lep~b} to within approximately \SI{1}{km \per s} using VLT/HiRISE \citep{vigan2024first}, while \citet{parker2024into} and \citet{landman2024betapicb} achieved measurements of $\upbeta$~Pic~b with uncertainties of approximately \SI{2}{km \per s} and \SI{300}{m \per s}, respectively. Using KPIC \citep{delorme2021keck}, \citet{horstman2024rv} was able to obtain RV epochs of \object{GQ~Lup~B} with precisions ranging from \SI{400}{} to \SI{1000}{m \per s}.
Notwithstanding these promising results that will stimulate further development of the involved techniques and instruments, the attainable planetary RV precision does not yet suffice to detect the putative P24-moon.

Regardless of whether one pursues an RV or astrometry-based strategy, a single epoch will never suffice for detection. Instead, extensive monitoring of the planet is necessary to reveal its periodic reflex motion.
Accordingly, the study of a simulated time series of mock epochs is the logical next step towards assessing what is required to astrometrically detect an exomoon.

\section{Modelling multi-epoch detection attempts}
\label{section_modelling_multi_epoch_detection_attempts}

To probe which kind of systems a given instrument is sensitive to, we need to establish different observation sampling strategies, implement a procedure of generating mock astrometric epochs, provide safeguards against over- or underestimating the signal of a given moon, and define a robust way of computing the significance for any combination of instrument, observing strategy and system configuration.
In the following, these points will be addressed individually.

\subsection{Defining observing strategies}
\label{section_defining_observing_sequences}

The chosen observational cadence is of consequence for the system configurations one is most sensitive to.
First, it is important to note that there is no one optimal strategy.
Instead, the approach one adopts would ideally be informed by the target and the period one suspects the moon to exhibit.
Assessing which sampling strategy is ideal for which system is a far-reaching problem by itself.
Avoiding aliasing and achieving an efficient sampling of the astrometric curve are key concerns that need to be addressed (see \citealt{madore2005nonuniform}).
Doing so is beyond the scope of this work, however. We shall instead confine our study to investigating how the attainable sensitivity of different instruments behaves as a function of the number of astrometric epochs.
Second, most targets are not observable throughout the entire year from most locations on Earth. The chosen observatory and target system position thus impose a second constraint on the observing strategy.
Finally, there might be special boundary conditions set by the operational processes and time allocation procedure of the chosen observatory. For instance, on Cerro Paranal, the VLTI observation windows are scheduled to coincide with the full moon and typically last for five days each.
Since at the moment there is little to no possibility of deviating from this fixed schedule we shall consider it a hard condition in the analysis to follow.
Bearing the above in mind, we define two basic strategies:
\begin{enumerate}
    \item 12 epochs: A sequence of evenly spaced observations with a cadence of one synodic month, that is approximately \SI{29.5}{days}. Once six epochs have been obtained, we pause for a period of six months during which the target is assumed to be unobservable, before observing for another six months in a row.
    \item 18 epochs: Same as above but including a second pause and third sequence of six epochs.
\end{enumerate}
Naturally, these approaches only represent a small subset of all possible strategies and are mainly geared towards investigating the sensitivity gain obtained from an additional set of six astrometric epochs.
Beyond the above approaches, one could adopt a higher-frequency tactic where an astrometric epoch is obtained during every night of several consecutive VLTI windows. Such a ``rapid-fire'' strategy would be especially suited to hunt for short-period moons. As mentioned above, however, we defer a thorough signal-processing-based identification of optimal sampling strategies for different system configurations to future works.

\subsection{Creating astrometric mock epochs}
\label{subsection_creating_astrometric_mock_epochs}
\begin{figure*}
    \centering
    \includegraphics[width=0.99\textwidth]{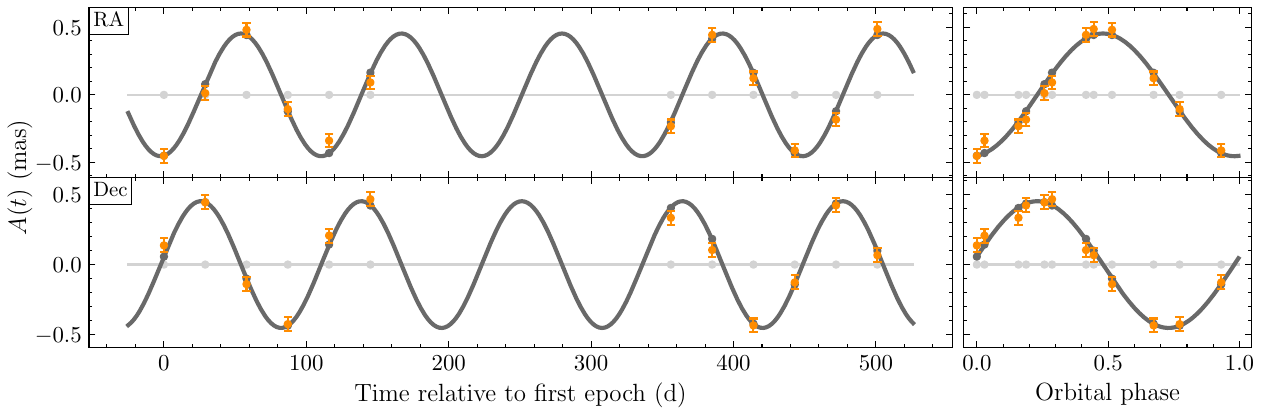}
    \caption{
    Fiducial astrometric signal in RA and Dec as a function of time for an exemplary system configuration ($M_\mathrm{s}=\SI{1.5}{M_\odot}$, $M_\mathrm{pl}=\SI{10}{M_{Jup}}$, $a_\mathrm{pl,\,0}=\SI{10}{AU}$, $M_\mathrm{m}=\SI{1}{M_{Jup}}$, $\varpi=\SI{50}{mas}$, $a_\mathrm{m,\,0}=\SI{0.1}{AU}$, $i_\mathrm{m}=\SI{0}{^\circ}$, $e_\mathrm{m}=\SI{0}{}$; see caption of Fig.~\ref{figure_wobble_schematic} for index definitions). The dark grey curve indicates the fiducial signal computed using the star--planet--moon model described in Sect.~\ref{subsubsection_star_planet_moon_model}, while the light grey line shows the zero-signal expected in the absence of a moon. The light and dark grey circles show the fiducial momentary signals in RA and Dec when employing the 12 epoch strategy defined in Sect.~\ref{section_defining_observing_sequences}. In orange we indicate the associated GRAVITY mock epochs generated according to the procedure outlined in Sect.~\ref{subsection_creating_astrometric_mock_epochs}. The right hand panels show the same data but phase-folded by the fiducial orbital period.
    }
    \label{figure_example_signal_sampling_in_ra_and_dec}
\end{figure*}

Based on the chosen number of epochs, we next need to generate a mock astrometric data set.
To this end, we can use the models introduced in Sect.~\ref{section_astrometric_signal} as our underlying fiducial models.
They provide the exact on-sky positions of the star, planet and moon relative to the system barycentre for any given time of observation. By extracting the planetary position according to the two models at a series of pre-defined observing times, we obtain our fiducial planet positions. These are not the positions that our instrument will measure, however. Shifting them in both RA and Dec by a noise component drawn from the Gaussian distribution $\mathcal{N}(0, \sigma_A)$, where $\sigma_A$ is the measurement uncertainty of the chosen instrument, turns the fiducial planet positions into our mock astrometric epochs.
Figure~\ref{figure_example_signal_sampling_in_ra_and_dec} shows the fiducial signal and generated GRAVITY mock epochs as a function of time in RA and Dec for an exemplary system configuration.

\subsection{Randomising the time of periplanet passage}
\label{subsection_time_of_periplanet_passage_randomisation}

The fact that the astrometric signal varies periodically over time holds the danger of significantly misjudging the achievable detection significance. In case the epochs are taken at regular intervals and the chosen cadence coincides approximately with a multiple of $P_\mathrm{m}/2$, where $P_\mathrm{m}$ is the orbital period of the moon, one runs the risk of consistently sampling the wobble at the same angular deviation. For instance, one could conceivably ``be lucky'' and systematically observe the planet at its largest deviation from the Keplerian trajectory or ``be unlucky'' and catch it at just the moments when the astrometric signal is minimal.
To avoid this pitfall when hunting for moons in earnest, it is therefore advisable to refrain from employing an evenly spaced observing strategy in cases where one expects the moon orbit to be of a similar period.
In our simulation, however, we can additionally counteract the issue by randomising the time of periplanet passage, $t_\mathrm{peri,m}$.
This effectively entails a randomisation of the planet--moon orbital phase such that an evenly spaced observing sequence can still coincidentally align with the planetary wobble for a single run but not for the entire sample of $N_\mathrm{rand}$ runs.
Another consequence of this manoeuvre is that for a given star--planet--moon configuration we will be left with a distribution of detection significances.

\subsection{Estimating the detection significance}
\label{subsection_computing_detection_significance}
The last ingredient required to assess our current and future detection capabilities is a framework that allows us to quantify in which parameter spaces the star--planet--moon model performs significantly better at explaining the mock astrometric epochs than the mere star--planet model.

The two models are nested in the sense that the star--planet--moon model transitions into the star--planet model when the moon mass becomes negligible. That being said, the star--planet--moon model requires more free parameters, which implies that any approach to be used to compare the performance of the two models must be capable of penalising such increased complexity.
Since it computes the Bayesian evidence, a nested sampling approach (see e.g.\,\citealt{feroz2008multimodal, buchner2021ultranest}) would enable us to account for the number of free parameters in a given model. As we are planning to compute the significance achievable by different instruments using various observing strategies over a grid of target system parameters, while also randomising the time of periplanet passage for each grid point (see Sect.~\ref{subsection_time_of_periplanet_passage_randomisation}), such a strategy would require inordinate computational effort.
Given the general character of this work as an initial feasibility assessment, we decided to defer the implementation of a nested sampling routine to a future study and focus the sensitivity estimates on comparing the $\chi$-squared values presented by the two models at the fiducial parameter settings.
This approach presumes that the fiducial solution can be converged upon using a suitable fitting routine, an assumption that is justified for the cases we are most interested in: the ones where $\Delta\chi^2_\mathrm{red}$, the difference between the reduced $\chi$-squared values of the two models, is significant.
Starting from the fiducial parameter set we ran a gradient descent routine to ensure the minimum was stable.
The $\chi$-squared values used throughout the remainder of the analysis presented here are based on these converged-upon parameter sets.

To quantify the relative performance of the two models, we applied a variant of the F-test based on the Fisher-Snedecor distribution (see e.g.\,\citealt{eadie1971statistical}).
Following \citet{band1997batse}, a given $\Delta\chi^2_\mathrm{red}$ value was converted to an $F$-value according to
\begin{equation}
    F = \frac{\Delta\chi^2/\Delta\nu}{\chi^2_\mathrm{moon}/\nu_\mathrm{moon}} ,
    \label{equation_f_test}
\end{equation}
where $\Delta\nu$ is the difference in the degrees of freedom $\nu = N_\mathrm{data} - N_\mathrm{fp} + 1$ between two models. Here, $N_\mathrm{data}$ and $N_\mathrm{fp}$ are the number of data points and free parameters, respectively. To account for the fact that each astrometric epoch corresponds to a measurement in RA and Dec, we used $N_\mathrm{data}=2N_\mathrm{epoch}$. The quantities in the denominator of Equation~\ref{equation_f_test} are the $\chi$-squared and degrees of freedom of the more complex model, that is the one including the moon.
From the resulting $F$-values, we computed the corresponding $p$-values via its survival function. Finally, we converted the $p$-values into a significance via the quantile function. Both operations were handled using the respective \texttt{scipy} methods \citep{virtanen2020scipy}.

We thus possess all the tools required to collapse the $\chi$-squared values resulting from the comparison of the two models with a set of noisy mock data obtained by a given instrument employing a certain observing sequence of a specific star--planet--moon configuration into a single detection significance.

\section{Multi-epoch sensitivity curves}
\label{section_sensitivity_curves}

We applied the procedure outlined above to estimate the detection significance achievable using VLTI/GRAVITY, VLTI/PLANETES and a hypothetical future interferometric facility with a \SI{3}{km} baseline (see Table~\ref{table_astrom_amplitude_and_precision_examples} for the respective planetometric measurement uncertainties), each employing the two observing strategies described in Sect.~\ref{section_defining_observing_sequences}.
To facilitate comparability with a large fraction of the directly imaged planet population, the fictional star--planet--moon system this investigation is based on was deliberately configured as general as possible.
Accordingly, we chose a \SI{10}{M_{Jup}} planet revolving around a \SI{1.5}{M_\odot} star on a circular, face-on orbit ($i= \SI{0}{^\circ}$) with a semi-major axis of \SI{10}{AU} and at a distance of \SI{20}{pc}.
The moon orbit was assumed circular and co-planar with the planetary orbit, that is also face-on.
The planetary Hill radius resulting from these fiducial parameter settings amounts to approximately \SI{1.2}{AU}.
These parameters remained fixed throughout the grid exploration of a two-dimensional plane spanned by the moon semi-major axis and moon mass axes, which were logarithmically sampled by \SI{15}{} values from \SI{e-2}{} to \SI{e0}{AU} and \SI{e-2}{} to \SI{e0}{M_{Jup}}, respectively.
Each grid point was sampled \SI{100}{} times, generating a new set of mock epochs with a randomly drawn noise term, respectively. This ensures adequate sampling of the underlying $\chi$-squared statistic and sufficient randomisation of the time of periplanet passage which was drawn anew for each sample (see Sect.~\ref{subsection_time_of_periplanet_passage_randomisation}).
Accordingly, this yielded a sampling of \SI{100}{} $\chi$-squared values per model fit performed.

\begin{figure*}
    \centering
    \includegraphics[width=0.99\textwidth]{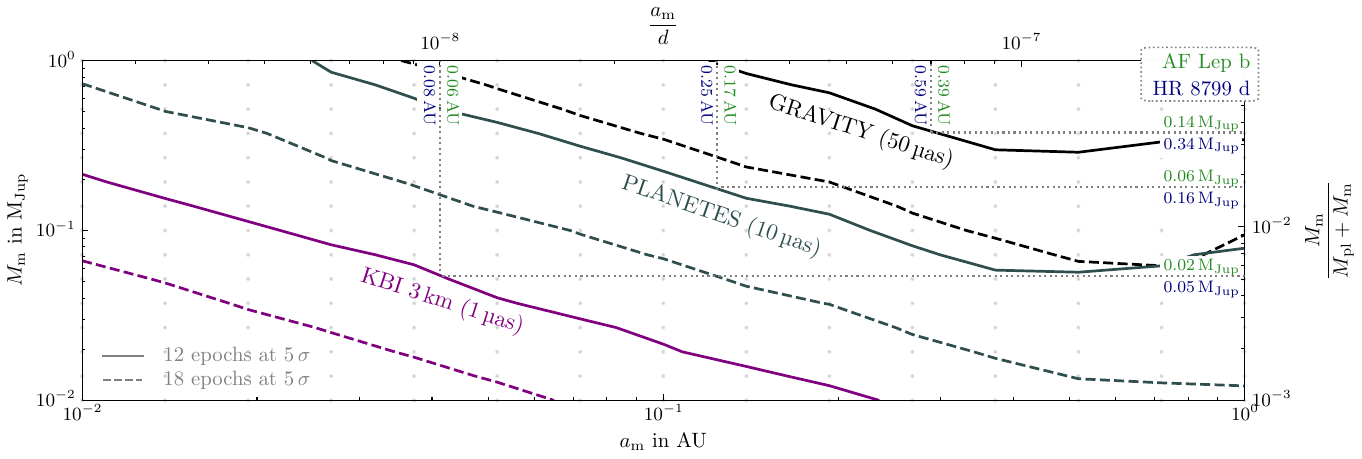}
    \caption{
    Sensitivity curves of different current and future interferometric instruments capable of astrometric measurements.
    The solid and dashed lines indicate the $\SI{5}{\sigma}$ detection limits based on 12 and 18 epochs, respectively.
    The grey dots in the background mark the probed grid points.
    The rescaled axes in the top and on the right provide a parametrisation where the fiducial target system distance, $d=\SI{20}{pc}$, and planet mass, $M_\mathrm{pl}=\SI{10}{M_{Jup}}$, have been folded in. They can thus be used to estimate the moon detection significance achievable for a genuine exoplanet of interest. Three exemplary moons detectable around ($M_\mathrm{pl}= \SI{3.27(25)}{M_{Jup}}$; \citealt{balmer2025aflepb}) and HR~8799~d ($M_\mathrm{pl}=\SI{9.3(5)}{M_{Jup}}$; \citealt{zurlo2022orbital}) are indicated by the dotted lines. The corresponding moon masses and semi-major axes are provided.
    }
    \label{figure_detection_significance_map}
\end{figure*}

At this point, one could choose between retaining the minimum, median or maximum $\chi$-squared value (and associated significance computed via the F-test routine) depending on whether one is interested in extracting the least favourable, median or most favourable detection scenario, respectively.
To arrive at a realistic assessment of our detection capabilities we opted to retain the median $\chi$-squared value for each model and grid point.
The $\SI{5}{\sigma}$ detection contours resulting from the different combinations of instruments and observing strategies are visualised in Fig.~\ref{figure_detection_significance_map}.

Besides the number of obtained epochs, the detection significance at a given grid point in Fig.~\ref{figure_detection_significance_map} is mainly governed by the expected astrometric signal amplitude.
In Equation~\ref{equation_astrometric_signal}, we found that $A \propto a_\mathrm{m}/d$ and $A \propto M_\mathrm{m}/(M_\mathrm{pl} + M_\mathrm{m})$.
By appropriately folding in the fiducial distance, $d$, and the total mass, $M_\mathrm{pl} + M_\mathrm{m}$, of the fictional underlying planet--moon system, we rescaled the axes in Fig.~\ref{figure_detection_significance_map} providing the resulting alternate axes along the right and top side of the panel. These rescaled axes enable comparison of the obtained sensitivity contours with genuine exoplanets that might be considered for a follow-up aimed at unveiling an exomoon.

To give two illustrative examples, we included the cases of AF~Lep~b ($M_\mathrm{pl}= \SI{3.27(25)}{M_{Jup}}$; \citealt{balmer2025aflepb}) and HR~8799~d ($M_\mathrm{pl}=\SI{9.3(5)}{M_{Jup}}$; \citealt{zurlo2022orbital}). Multiplying the values along the rescaled semi-major axis axis with the distance of a target system yields the physical semi-major axis values corresponding to the significance contours. Likewise, using the mass of an exemplary target planet, one can convert the values along the mass ratio axis into the corresponding moon masses.
Doing so, we find that 12 astrometric GRAVITY epochs suffice to detect a moon of \SI{0.34}{M_{Jup}} at a separation of \SI{0.59}{AU} around HR~8799~d with a confidence of $\SI{5}{\sigma}$.
Targeting the lower mass AF~Lep~b, 12 epochs could reveal a \SI{0.14}{M_{Jup}} moon at a separation of \SI{0.39}{AU}. Both of these hypothetical moons would reside within $0.5R_\mathrm{Hill,\,pl}$ of their respective planetary hosts.
Since more massive or more distant moons would induce a larger astrometric signal, they would manifest themselves in the data at even higher significances. Conversely, if none is present, 12 epochs would be adequate to rule out the existence of such moons around AF~Lep~b and HR~8799~d.
The arrival of PLANETES at the VLTI will amplify our sensitivity such that moons of \SI{0.06}{M_{Jup}} at \SI{0.17}{AU} and \SI{0.16}{M_{Jup}} at \SI{0.25}{AU} were detectable at $\SI{5}{\sigma}$ around AF~Lep~b and HR~8799~d, respectively.
Finally, a future \SI{3}{km} baseline interferometric facility will be capable of detecting moons of \SI{0.02}{M_{Jup}} at \SI{0.06}{AU} and \SI{0.05}{M_{Jup}} at \SI{0.08}{AU} around the two planets at the same confidence.
These scenarios, indicated by the grey dotted lines in Fig.~\ref{figure_detection_significance_map}, are only three example cases for two arbitrary exoplanets. 
Note that \SI{0.02}{} and \SI{0.05}{M_{Jup}} translate to approximately \SI{6}{} and \SI{16}{M_{Earth}}, respectively.
Thus, the moons detectable with a \SI{3}{km} baseline interferometric facility are comparable with Super-Earth exoplanets \citep{valencia2007radius}.
At higher moon masses the instruments would be capable of probing smaller semi-major axes. Similarly, at larger moon semi-major axes the detectable masses can be substantially lower.
Targeting potential host planets of lesser mass than either AF~Lep~b or HR~8799~d in their respective circumstellar habitable zones may eventually teach us how prevalent potentially habitable exomoons are.
The recently imaged planet candidate around $\alpha$~Cen~A \citep{beichman2025worlds, sanghi2025worlds} might provide a suitable first target as it combines the favourable characteristics of being nearby and low in mass while at the same time orbiting within the circumstellar habitable zone.

As long as they are comparable to the underlying fiducial system configuration on which the significance contours are based (far-out, circular, face-on orbit planets), Fig.~\ref{figure_detection_significance_map} can be used to estimate the attainable instrument-dependent detection significance for any planet of interest.
Star--planet--moon configurations and viewing angles that differ from the underlying fiducial system such as eccentric or edge-on orbits warrant dedicated simulations.
Intuitively, for edge-on or near-edge-on cases the signal amplitude would remain the same, even if the reflex motion of the planet were confined to tracing out a line as projected onto the plane of the sky.
With the publication of the astrometric time series data in \textit{Gaia} DR4, we can expect to confirm dozens of planet candidates using GRAVITY. The combination of \textit{Gaia} astrometry and the direct detection will yield a large sample of narrowly constrained orbital solutions and precise dynamical masses (see \citealt{winterhalder2024combining}) that will lay the groundwork for an astrometric exomoon hunt.

Broadly speaking, based on 12 or 18 epochs, GRAVITY and PLANETES are sensitive at $\SI{5}{\sigma}$ to moons between \SI{0.1}{} and a few per cent of the mass of their planetary host.
According to \citet{canup2006common}, moons that have formed in-situ, that is in orbit around the planetary host, should not exceed a moon-to-planet mass ratio of approximately $\SI{e-4}{}$.
There thus does not exist an overlap between our current and near-future astrometric exomoon detection capabilities and the regime where these regular moons are expected to reside.
Judging by Fig.~\ref{figure_detection_significance_map}, a kilometre-baseline facility might begin to tackle this limit -- at least for comparatively large moon semi-major axes.
So-called irregular moons, that is ones that were either captured by or formed as a result of an impact with their host, do not abide by this upper mass-ratio limit.

To assess the potential of an astrometric search for exomoons, it is of interest what expenditure in observation time such a study would entail.
Based on experiences drawn from observations conducted within the framework of the ExoGRAVITY Large Programme \citep[ESO ID 1104.C-0651][]{Lacour_ExoGRAVITY_LP}, obtaining one GRAVITY epoch can be assumed to require a total of 2 hours of telescope time including overheads. Data sets comprising 12 and 18 epochs, such as the ones the sensitivity curves in Fig.~\ref{figure_detection_significance_map} are based on, would therefore carry a cost between 20 and 40 hours of VLTI UT-mode time. While this is a non-negligible investment, a potential moon discovery would make such a project highly worthwhile.
Risk mitigation can be achieved by an informed target selection that hinges on clues as to the potential existence of an orbiting moon from auxiliary studies.
The spectroscopic data products that result from a GRAVITY observation can act as contingency science cases in the event that no moon is present or detectable: time-series observations of exoplanet and brown dwarfs can be probed for variability or combined into a single high signal-to-noise spectrum (see the ExoGRAVITY spectral library to be presented in Kammerer et al., in prep.).

Finally, a note on distinguishing between perturbations caused by a planet-orbiting moon and a star-orbiting interior planet, since, in general, both object classes would manifest themselves as an astrometric wobble. The expected periods are significantly different, however. Whereas the sought-after moons are expected to impart signals with periods of days to months, inner planets can be assumed to complete their orbits on typical timescales of several years \citep{lacour2021mass}. There does exist an overlap between the two cases, however: undetected hot Jupiters would induce a stellar reflex motion that could mimic the signature of a moon since our astrometric epochs correspond to relative separation measurements between the host star and the targeted planet. Such niche cases could easily be handled by obtaining an RV data set of the star. If present, a signal of a similar period would strongly suggest an inner planet causing the observed astrometric wobble. An exomoon, on the other hand, would not impart any measurable stellar RV modulation.

\section{Conclusions}
\label{section_conclusions}

In this feasibility study, we have demonstrated the potential of hunting for exomoons and binary planets by means of an astrometric time series.
The unprecedented precision of interferometric instruments such as VLTI/GRAVITY renders the technique viable and sensitive to low-mass moons around far-out directly imaged planets, a regime where moons are expected to survive over long periods.
While, contrary to the RV method, the astrometric signal drops for target systems at greater distances, if successfully applied, the method can provide constraints on all orbital parameters including the absolute dynamical mass of the moon, the key parameter upon which any characterisation of the satellite object depends.
The first dedicated GRAVITY search for an exomoon targeting \object{HD~206893~B} is currently being conducted with the results expected to be published soon (Kral et al. in prep).
We showed that, a next-generation instrument like VLTI/PLANETES will likely provide an astrometric measurement precision comparable to the signature induced by the putative exomoon around $\upbeta$~Pic~b as suggested by \citet{poon2024potential}.

The sensitivity curves obtained from modelling the astrometric signatures expected for different target system configurations quantify our current and future capabilities and provide a compelling science case for pursuing kilometre-baseline interferometric facilities.
Future studies should concentrate on implementing a fitting framework based on a nested sampling routine in order to robustly assess the degree to which a genuine astrometric time-series suggests the existence of an exomoon around a given planet. Moreover, the signal-processing conundrum of how to optimally sample an astrometric signature of a given period should be examined. In particular, such an investigation would help answer the question of whether the strict monthly VLTI observation window allocation inhibits the detection of short-period signals and whether the current scheduling system should be reconsidered.

On the whole, astrometric time-series observations provide an excellent means of detecting moons around gas giant exoplanets.
Intriguingly, a future kilometre-baseline interferometric facility, used for monitoring gas giants that orbit within their respective habitable zones, will be capable of detecting and characterising exomoons with masses comparable to that of Earth.

\begin{acknowledgements}
SL acknowledges funding from the European Union (ERC AdG 101142746 PLANETES). Views and opinions expressed are however those of the author(s) only and do not necessarily reflect those of the European Union or the European Research Council. Neither the European Union nor the granting authority can be held responsible for them.
\end{acknowledgements}

\bibliographystyle{aa_url.bst}
\bibliography{refs.bib}

\end{document}